\documentstyle[11pt,psfig]{article}
\def\bi{\bibitem{}}

\def\ni{\noindent}
\def\beb{\begin{thebibliography}{999}}
\def\eeb{\end{thebibliography}}
\def\bei{\begin{itemize}}
\def\eei{\end{itemize}}
\def\bef{\begin{figure}}
\def\eef{\end{figure}}
\def\ben{\begin{enumerate}}
\def\een{\end{enumerate}}
\def\beq{\begin{equation}}
\def\eeq{\end{equation}}
\def\ber{\begin{eqnarray}}
\def\eer{\end{eqnarray}}

\newcommand{\dmdt}{{\mbox{{\rm M}$_{\odot}$}} {\rm yr}$^{-1}$}

\newcommand{\lsim}{\raisebox{-0.3ex}{\mbox{$\stackrel{<}{_\sim} \,$}}}
\newcommand{\gsim}{\raisebox{-0.3ex}{\mbox{$\stackrel{>}{_\sim} \,$}}}
\begin{document}
\centerline{\Large Whither Strange Pulsars?}
\centerline{Sushan Konar} 
\centerline{\small {Inter-University Centre for Astronomy and Astrophysics, Pune 41107, India}}
\baselineskip=12pt

\begin{abstract}
It has been argued that some of the pulsars could be strange stars because these too are capable of supporting 
fast rotations observed in pulsars. We examine this claim from the point of view of the evolution of magnetic 
field in pulsars and find that there is no compelling reason for the existence of strange pulsars.
\end{abstract}

\ni {\em Key Words} : magnetic fields, stars--neutron, stars--strange, pulsars-general \\

\section{Introduction}
\ni {\em Strange Quark Matter} (SQM), composed of u, d and s quarks, may probably be the ultimate ground state of 
matter (Farhi \& Jaffe 1984, Witten 1984). If meta-stable at zero pressure it might exist in the central 
region of compact objects stabilized by the high pressure (Glendenning, Kettner \& Weber 1995). If however, SQM is 
absolutely stable at zero pressure the existence of {\em Strange Stars} is a possibility. The stable range of mass 
and radius of strange stars are similar to neutron stars, hence the claim that some or all the pulsars are strange 
stars (Alcock, Farhi \& Olinto 1986). In this work, we review the {\em strange pulsar} hypothesis from the point of 
view of the evolution of the magnetic field and investigate : a) the maximum field strength sustainable by strange stars,
b) the possible current configurations supporting the field, and c) the evolution of such fields in isolated as well as 
accreting strange stars. We compare our results with the known observational facts concerning the nature of the pulsar 
magnetic fields.

\section{Maximum Field Strength}
\ni For fields larger than $\sim 4.4 \times 10^{13}$~G in the core of a proto-neutron star the deconfinement transition 
is strongly suppressed preventing the formation of a strange star in an SNE with a higher field (Ghosh \& Chakrabarty 1998). 
On the other hand, for a strange star forming via the deconfinement conversion of an accreting neutron star (Olinto 1987) 
the maximum field would be that applicable for the neutron stars ($\sim 10^{15}$~G). For a strange star with a hadronic 
crust, though, the maximum field is determined by the shear stress of the crust ($\sigma_{\rm shear} \gsim P_{\rm mag}$)
and is found to be $ \sim 5 \times 10^{13} \mbox{G}$. Though sufficient for ordinary pulsars - it falls short of the field 
strength of exotic objects like magnetars.

\section{Field Configuration}
\ni According to the recent models, strange stars have two distinct regions - a quark core and a thin crystalline nuclear 
crust separated by a dipole layer of electrons (Glendenning \& Weber 1992). Since no currents can flow across a dipole 
layer the currents, supporting a magnetic field, either reside entirely within the quark core or are completely confined 
to the nuclear crust. The evolution of the magnetic field is governed by the equation :
\beq
\frac{\partial \vec B}{\partial t} = - \frac{c^2}{4 \pi} 
\nabla \times (\frac{1}{\sigma} \times \vec \nabla \times \vec B),
\label{e_dbdt}
\eeq
where $\sigma$ is the electrical conductivity of the medium. The ohmic dissipation time-scale is, 
$\tau_{\rm ohmic} \sim \frac{4 \pi}{c^2} \sigma L^2$, where $L$ is the system dimension. If the currents are 
in the crust then for allowable range of crustal parameters we find that the dissipation time-scale is 
$\lsim 3 \times 10^6 \mbox{years}$. This is much smaller than the typical time-scale in which the field remains
stable in millisecond ($\sim 10^9$ yrs) as well as in isolated pulsars ($\sim 10^8$ yrs). Hence, a crustal field
is not long-lasting enough for the strange star to effectively function as a pulsar.

\section{field evolution}
\ni Assuming the quarks to be massless in the u,d,s plasma in the core, the conductivity is (H\t{ae}nsel \& Jerzak 1989)
$\sigma \sim 6 \times 10^{25} (\frac{\alpha_c}{0.1})^{-3/2} T_{10}^{-2} \frac{n_B}{n_{B0}} s^{-1}$, where $\alpha_c$ is 
the QCD coupling constant, $T_{10}$ is temperature of the isothermal core in units of $10^{10}$~K, $n_B$ is the number 
density of baryons and $n_{B0}$ is the nuclear baryon number density. The main effect of accretion is to raise the 
temperature of the star. Assuming the strange star thermal evolution to be at least as fast as or faster than that of 
the neutron stars, the temperature of an accreting strange star should be equal to or less than that of an accreting 
neutron star. This gives a lower limit to the dissipation time-scale. Using $T \sim 10^9$~K corresponding to an accretion 
rate of $\sim 10^{-9}$~\dmdt (Miralda-Escude et al. 1990) we get $\tau_{\rm ohmic} \sim 10^{13}$ years. This completely 
rules out the possibility of any field reduction even in an accreting strange star. \\

\ni Therefore, even if strange pulsars exist their magnetic field would not decay in an accreting system, contrary to 
the expectation from extensive pulsar observation. Hence, there is as yet no compelling arguments in favour of strange 
pulsars vis-a-vis neutron stars to function as pulsars.  

\section*{Acknowledgment}
Discussions with Dipankar Bhattacharya, Bhaskar Datta, Jes Madsen and Fridolin Weber have been very helpful.

\beb
\baselineskip=10pt
\bi{} Alcock C., Farhi E., Olinto A., 1986, ApJ, 310, 261
\bi{} Farhi E., Jaffe R.~L., 1984, PRD, 30, 2379 
\bi{} Glendenning N.~K., Weber F., 1992, ApJ, 400, 647
\bi{} Glendenning N.~K., Kettner Ch., Weber F., 1995, ApJ, 450, 253
\bi{} Ghosh T., Chakrabarty S., 1998, astro-ph/9811243
\bi{} H\t{ae}nsel P., Jerzak A.~J., 1989, Acta Physica Polonica B, 20, 141
\bi{} Miralda-Escude J., H\t{ae}nsel P., Paczynski B., 1990, ApJ, 362, 572
\bi{} Olinto A.~V., 1987, PLB, 192, 71
\bi{} Witten E., 1984, PRD, 30, 272 
\eeb

\end{document}